\documentclass[aps,prl,twocolumn,nopacs,superscriptaddress]{revtex4}
\usepackage{graphicx}
\usepackage{verbatim}
\usepackage{amsmath}
\usepackage{sidecap}
\usepackage{epstopdf}
\usepackage{braket}

\begin{document}

\title{Measurement of collective excitations in VO$_2$ by resonant inelastic X-ray scattering}

%\title{Measurement of charge and spin collective excitations in VO$_2$ at the energy scale of the insulating band gap}
%\title{Measurement of charge and spin collective excitations in VO$_2$ across the metal-insulator transition}
%\title{Observation of electronic collective excitations in VO$_2$ across the metal-insulator transition}
%\title{Collective excitations of VO$_2$ measured across the metal-insulator transition with resonant inelastic X-ray scattering}
%\title{Strongly correlated energy levels of VO$_2$ measured across the metal-insulator transition with resonant inelastic X-ray scattering}
%\title{Excitation spectra of VO$_2 dimers and chains measured with resonant inelastic X-ray scattering}

\author{Haowei He}
\affiliation{Department of Physics, New York University, New York, New York 10003, USA}
\author{A. X. Gray}
\email{axgray@temple.edu}
\thanks{Corresponding author}
\affiliation{Stanford Institute for Materials and Energy Sciences, SLAC National Accelerator Laboratory, 2575 Sand Hill Road, Menlo Park, California 94025, USA}
\affiliation{Department of Physics, Temple University, 1925 N. 12th St., Philadelphia, Pennsylvania 19130, USA}
%\author{Yishuai Xu}
%\author{Janet Chiu}
%\affiliation{Department of Physics, New York University, New York, New York 10003, USA}
\author{P. Granitzka}
\affiliation{Stanford Institute for Materials and Energy Sciences, SLAC National Accelerator Laboratory, 2575 Sand Hill Road, Menlo Park, California 94025, USA}
\affiliation{Van der Waals-Zeeman Institute, University of Amsterdam, 1018XE Amsterdam, The Netherlands}
\author{J. W. Jeong}
\author{N. P. Aetukuri}
\affiliation{IBM Almaden Research Center, San Jose, CA 95120, USA}
\author{R. Kukreja}
\affiliation{Stanford Institute for Materials and Energy Sciences, SLAC National Accelerator Laboratory, 2575 Sand Hill Road, Menlo Park, California 94025, USA}
\author{Lin Miao}
\affiliation{Department of Physics, New York University, New York, New York 10003, USA}
\affiliation{Advanced Light Source, Lawrence Berkeley National Laboratory, Berkeley, CA 94720, USA}
\author{Y. B. Huang}
\affiliation{Research Department Synchrotron Radiation and Nanotechnology, Paul Scherrer Institut, CH-5232 Villigen PSI, Switzerland}
\affiliation{Beijing National Laboratory for Condensed Matter Physics, and Institute of Physics, Chinese Academy of Sciences, Beijing 100190, China}
\author{P. Olalde-Velasco}
\author{J. Pelliciari}
\affiliation{Research Department Synchrotron Radiation and Nanotechnology, Paul Scherrer Institut, CH-5232 Villigen PSI, Switzerland}
\author{W. F. Schlotter}
\affiliation{Linac Coherent Light Source, SLAC National Accelerator Laboratory, Menlo Park, California 94025, USA}
%\author{P. Shafer}
\author{E. Arenholz}
\affiliation{Advanced Light Source, Lawrence Berkeley National Laboratory, Berkeley, CA 94720, USA}
\author{T. Schmitt}
\affiliation{Research Department Synchrotron Radiation and Nanotechnology, Paul Scherrer Institut, CH-5232 Villigen PSI, Switzerland}
\author{M. G. Samant}
\author{S. S. P. Parkin}
\affiliation{IBM Almaden Research Center, San Jose, CA 95120, USA}
\author{H. A. D\"{u}rr}
%\email{hdurr@slac.stanford.edu}
\affiliation{Stanford Institute for Materials and Energy Sciences, SLAC National Accelerator Laboratory, 2575 Sand Hill Road, Menlo Park, California 94025, USA}
\author{L. Andrew Wray}
\email{lawray@nyu.edu}
\thanks{Corresponding author}
\affiliation{Department of Physics, New York University, New York, New York 10003, USA}

\begin{abstract}

Vanadium dioxide is of broad interest as a spin-1/2 electron system that realizes a metal-insulator transition near room temperature, due to a combination of strongly correlated and itinerant electron physics. Here, resonant inelastic X-ray scattering is used to measure the excitation spectrum of charge, spin, and lattice degrees of freedom at the vanadium L-edge under different polarization and temperature conditions. These spectra reveal the evolution of energetics across the metal-insulator transition, including the low temperature appearance of a strong candidate for the singlet-triplet excitation of a vanadium dimer.

%found in the insulating state is likely to represent the lowest energy state split by the dimer.

%Comparison with a cluster calculation reveals the symmetries within the observed energy landshape, , and that an in-gap excitation found in the insulating state is likely to represent a singlet-triplet transition that breaks the vanadium dimer.

% (T$_{MIT}$ =340K)
%Vanadium dioxide is of broad interest as a spin-1/2 electron system that realizes a metal-insulator transition near room temperature (T$_{MIT}$ =340K). In a Mott picture, the transition is motivated by the formation of local singlet states, while in an itinerant picture it is motivated by the splitting of bands that are symmetric and antisymmetric with respect to dimerized vanadium sites. Here, resonant inelastic X-ray scattering is used to measure the electronic excitation spectrum at the vanadium L-edge under different polarization and temperature conditions. Comparison with theory reveals that the excitation features match expectations for a Mott ground state scenario, and that an in-gap excitation found in the insulating state may represent a dimer-breaking singlet-triplet excitation.

\end{abstract}
\maketitle

Vanadium dioxide is a spin-1/2 electron system that undergoes a metal-insulator transition near room temperature \cite{1}, and has been the subject of strong interest in both basic and applied research. When cooling through the transition, vanadium atoms pair into strongly hybridized dimers as the the crystal structure changes from rutile (R phase) to monclinic (M1 phase) \cite{2,3}. The mechanism driving this transition incorporates Peierls splitting of the bonding and antibonding states of the dimer basis \cite{6,7} and represents a fascinating crossover from itinerant to localized behavior in an electron system that is intrinsically poised at the threshold of becoming a Mott insulator \cite{12,13,14,15,KotliarDMFT2015,16}. A key challenge to establishing a comprehensive understanding of VO$_2$ based systems is that, though the gapping of symmetric and antisymmetric states within vanadium dimers is of central importance in motivating the metal to insulator transition, excitations across these gaps have not yet been experimentally resolved. Here, resonant inelastic X-ray scattering (RIXS) at the vanadium L-edge is used to measure the evolution of vanadium site energetics across the transition. Close comparison with a first principles based multiplet cluster model is used to identify symmetries within the RIXS spectrum, and reveals a strong candidate for a symmetric-to-antisymmetric excitation that breaks the singlet bond of a low temperature vanadium dimer.

%High resolution X-ray absorption spectroscopy (XAS) and RIXS at the vanadium L-edge ($2p$ core resonance) are performed to obtain spectra that represent unoccupied electronic states accessible through localized perturbations on the vanadium site. To interpret these results, a vanadium dimer Hamiltonian is diagonalized to solve the Kramers-Heisenberg scattering equation, incorporating first principles atomic multiplet and interatomic hybridization energetics.

%(T$_c$$=$298K)

\begin{figure}
\centering
\includegraphics[width = 8cm]{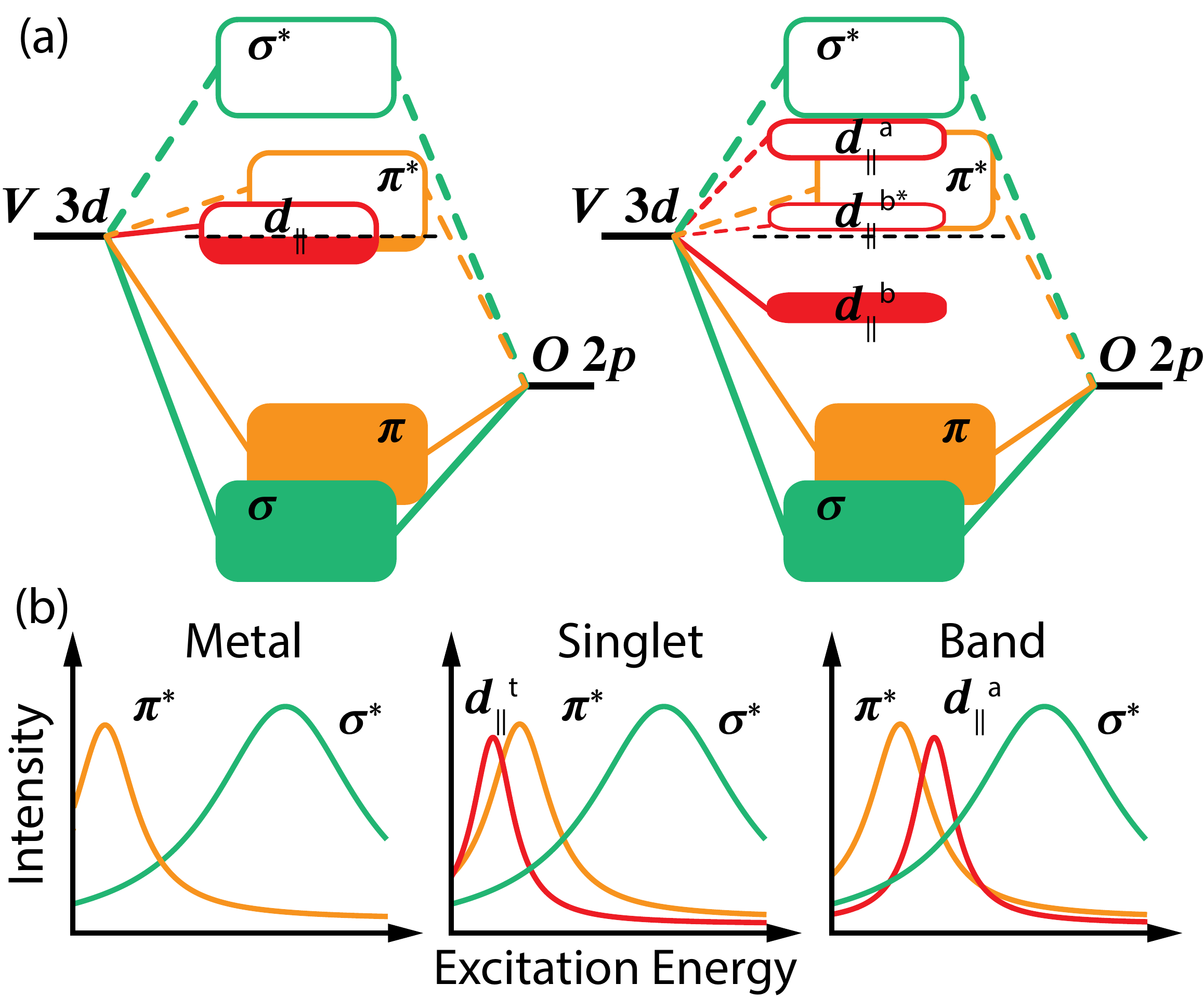}
\caption{{\bf{Orbital symmetries of VO$_2$ electronic states}}: (a) Orbital symmetries of electronic states are labeled for (left) the high-temperature metallic state and (right) the low-temperature insulating state, with the Fermi level indicated by a dashed line. The labels $d_{\parallel}^{b*}$ and $d_{\parallel}^{a}$ indicate gapped $d_{\parallel}$-symmetry states expected in the M1 phase. (b) Schematics show anticipated RIXS excitations that could be achieved by changing the symmetry of a $d_{\parallel}$ electron. Scenarios are presented for (left) the rutile metal and for hypothetical M1 insulating states based on (middle) strongly correlated singlets and (right) itinerant Peierls band physics. Higher energy excitations involving the oxygen $\pi$/$\sigma$ orbitals are not depicted.}
\label{fig:energyLevel}
\end{figure}

The orbital character of VO$_2$ electronic states was first explored by Goodenough \cite{19}, and is outlined in Fig. 1(a). The octahedral-like crystal field splits vanadium 3d orbitals into $\pi^*$ ($t_{2g}$) and $\sigma^*$ ($e_{g}$) manifolds. Electrons in the ground state largely occupy a $\pi^*$ orbital termed `$d_{\parallel}$', with lobes that point between neighboring vanadium atoms along the chain axis ($c_R$-axis). When cooling into the low temperature M1 phase (Fig.1(b)), dimerization of vanadium atoms along the $c_R$-axis splits the $d_{\parallel}$-derived states into symmetric and antisymmetric manifolds. The $d_{\parallel}$ derived states are split by the Peierls transition into bonding and antibonding states ($d_{\parallel}^b$ and $d_{\parallel}^a$), which are energetically modified and manifest additional DOS features (e.g. $d_{\parallel}^{b*}$) due to local entanglement and correlations \cite{15,KotliarDMFT2015,OxygenKdbstar}.

%Depending on whether one adopts an itinerant or strongly correlated perspective, the $d_{\parallel}$ density of states may primarily manifest as a single band strongly split into bonding and antibonding states by Peierls physics ($d_{\parallel}^b$ and $d_{\parallel}^a$), or as a wider set of features derived from these symmetries 

% 2-electron spin-split singlet and triplet spin states with a smaller gap of $\lesssim0.5eV$ ($d_{\parallel}^s$ and $d_{\parallel}^t$) \cite{15,KotliarDMFT2015}. 

% ?s?, ?t?, and ?*? to: ?b?, ?b*? and ?a?, 
%Because direct bonding-antibonding ($d_{\parallel}^b$ and $d_{\parallel}^a$) symmetry changing excitations of the dimer are nearly forbidden in the optical ($Q=0$) channel, 

%***need to add the oxygen O-edge study showing $d_{\parallel}^{b*}$

Elements of strong correlation have been well established in the electronic structure of VO$_2$ \cite{15,16,17,18,AlexCorrTransition}, and singlet bonding between dimerized vanadium sites with well defined $3d^1$ occupation is thought to explain the nonmagnetic nature of the insulating M1 phase. However, the excitations that best represent energetic changes upon entering the M1 phase, including bonding-antibonding ($d_{\parallel}^b$ to $d_{\parallel}^a$) excitations and the singlet-triplet excitation of vanadium dimers, fall within an antisymmetric sector that is not strongly accessed by optical ($Q\sim 0$) spectroscopies. Non-optical experimental studies of the electronic density of states have largely made use of single-particle spectroscopies in which an electron is added to or removed from a vanadium site, which does not give information about coherent electronic transitions such as the singlet-triplet mode. In the present study, soft X-rays provide sufficient momentum transfer along the dimer axis for direct symmetric-to-antisymmetric excitations to appear in RIXS spectra (see Fig. 3 discussion), making it possible to measure antisymmetric-sector excitations directly as outlined in Fig. 1(b).

%, and is most easily created by a spectroscopy in which electronic symmetry is manipulated coherently on a single site, as through core hole resonance.

\begin{figure}
\centering
\includegraphics[width = 8cm]{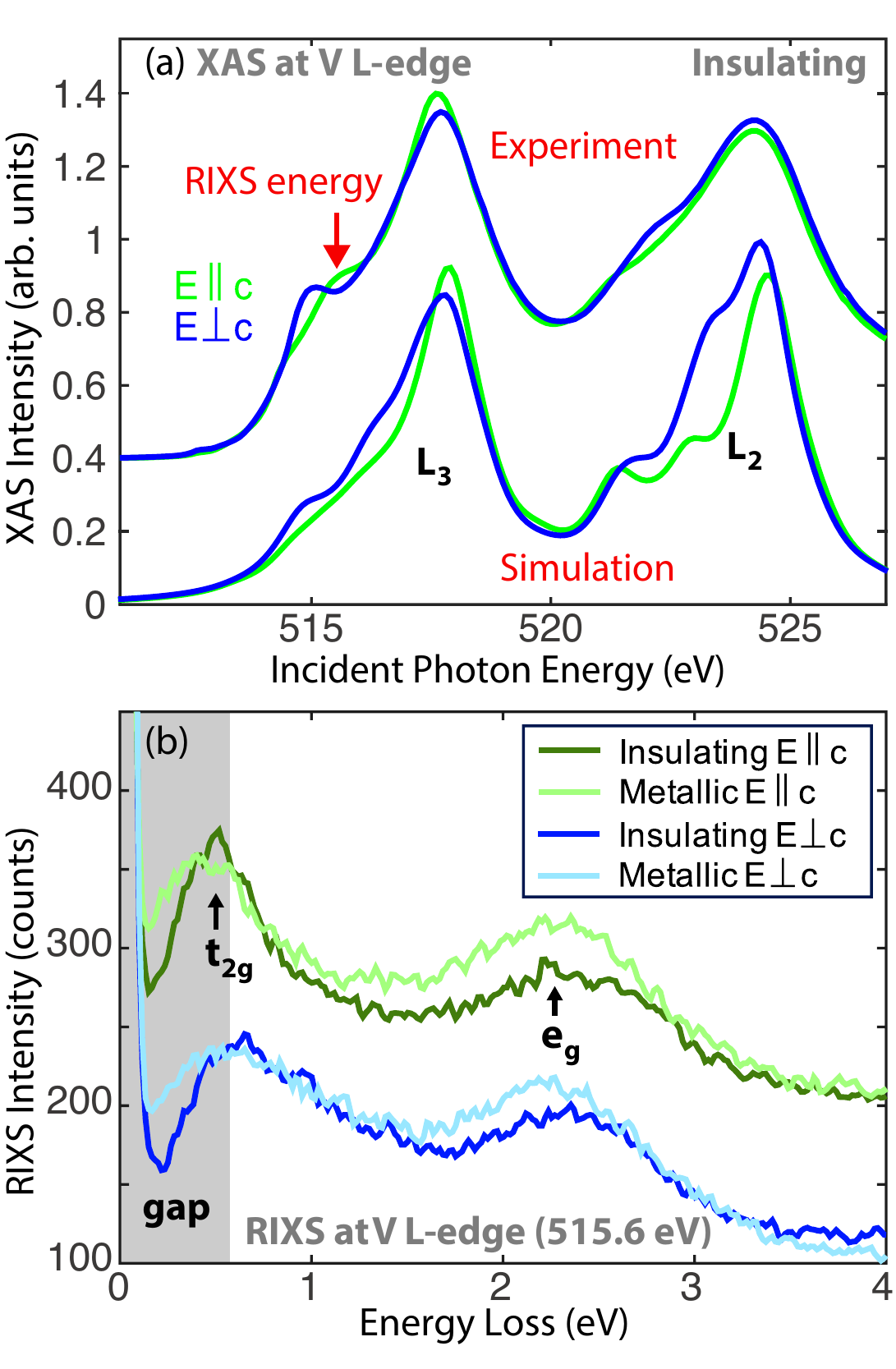}
\caption{{\bf{XAS and RIXS across the metal-insulator transition}}: (a) (top) Experimental and (bottom) theoretical XAS spectra of M1 phase VO$_2$ are shown for two incident photon polarization conditions. (b) RIXS spectra in the insulating and metallic states at temperatures T=260K and 320K, respectively. The two $\textbf{E} \perp \textbf{c}_R$ curves are downward offset by 100 counts.}
\label{fig:experiment}
\end{figure}

RIXS measurements were performed at the ADRESS beamline of the Swiss Light Source at the Paul Scherrer Institute \cite{ADRESS1,ADRESS2}, with combined energy resolution better than $\Delta$E=90 meV at the RIXS incident energy of $h\nu$=515.6 eV. The linear polarization of the incident photons could be set either perpendicular ($\sigma$-polarization) or parallel ($\pi$-polarization) to the scattering plane. This experimental configuration allows one to selectively probe excitations with polarization perpendicular ($\textbf{E} \perp \textbf{c}_R$) or near-parallel ($\textbf{E} \parallel \textbf{c}_R$) to the rutile $c_R$-axis. Measurements were carried out at temperatures of (insulating) 260 and (metallic) 320K, with base pressures better than 5$\times$10$^{-11}$ torr. The incident photons were maintained at a grazing angle of 15$^\circ$ and RIXS was measured at an acute outgoing angle of 65$^\circ$ with respect to the [001] sample surface (included angle is 50$^\circ$). Beam damage was minimized by adopting a new beam spot for each measurement. The high quality single crystalline VO$_2$ film of 10nm thickness was grown on a TiO$_2$(001) substrate by pulsed laser deposition, following the procedures described in Ref. \cite{sampleDesign}. Under these conditions, the metal-insulator transition occurs sharply at T$_{MI}$=295K, and the $c_R/a_R$ lattice constant ratio is 0.617 \cite{sampleDesign,AlexCorrTransition}.

The RIXS and XAS spectral functions are simulated for the experimental scattering geometry by the standard atomic multiplet method, augmented to incorporate two equivalently treated vanadium atoms with interatomic hopping via the $d_{\parallel}$ orbital (see details in online Supplemental Material \cite{SM}).
First principles calculations estimate the intra-dimer $d_{\parallel}$ hopping parameter to be more than an order of magnitude larger than inter-dimer $d_{\parallel}$ hopping \cite{15}, making this a good approximated basis for the analysis of low energy excitations in the M1 insulating state.

%The dimer Hamiltonian has eigenstates with momentum symmetries of Q = 0 and Q = $\pi/a$, where $a = \sqrt{2}\times1.9$\AA \hspace{0.1cm} is the separation between V-V pairs. From scattering theory, one can get the momentum (Q) transferred from the photon to the sample projects 84\% onto the Q = 0 component and  16\% onto the Q = $\pi/a$ component.

\begin{figure*}
\centering
\includegraphics[width = 16cm]{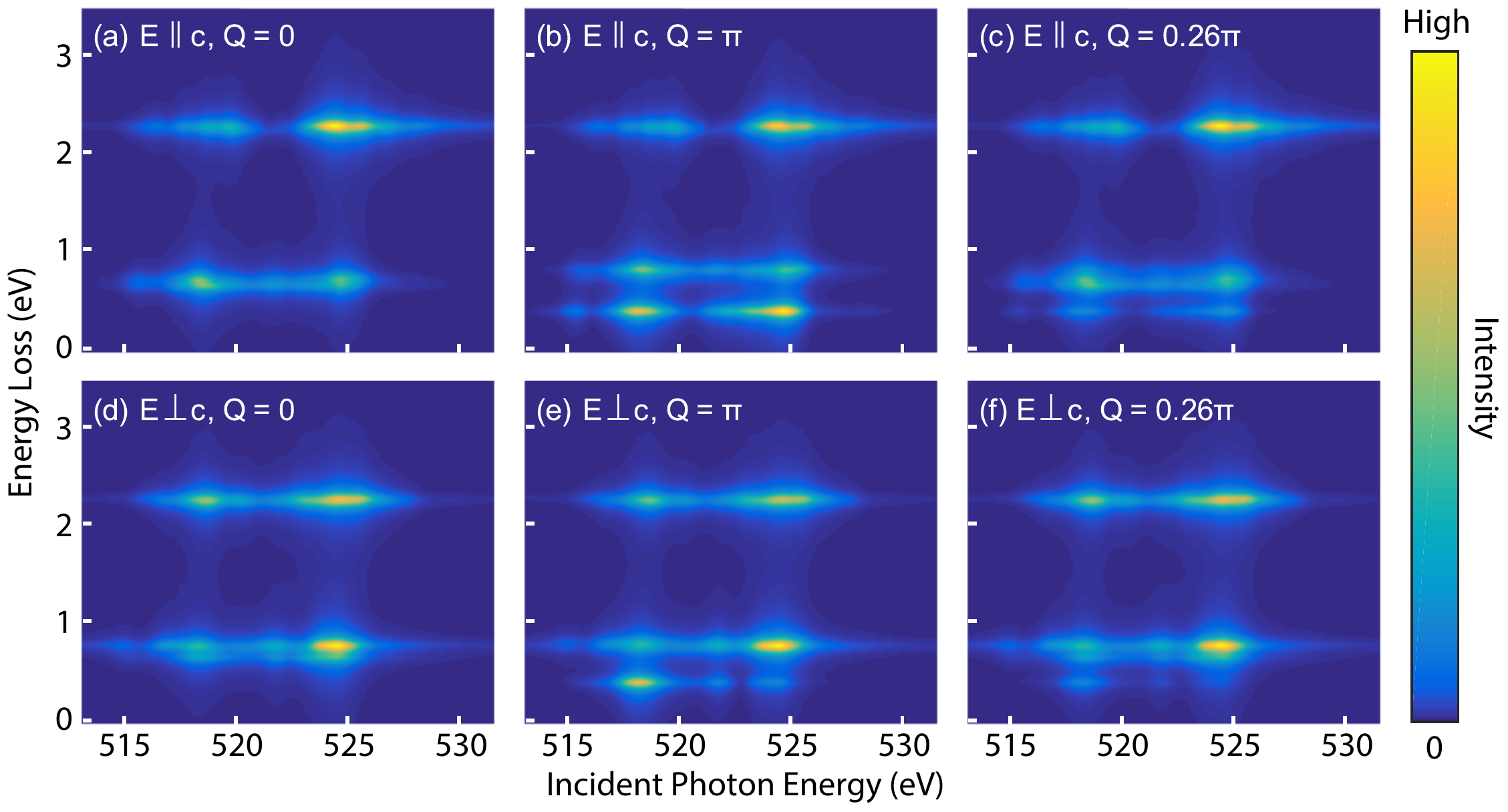}
\caption{{\bf{Simulated RIXS spectra of a vanadium dimer}}: The RIXS scattering profile is simulated for $\textbf{E} \parallel \textbf{c}_R$ polarized incident photons in the high symmetry (a) Q=0 and (b) Q=$\pi$ sectors, and (c) for a superposition given by the experimental value of Q=0.26$\pi$. Panels (d-f) show corresponding spectra obtained with $\textbf{E} \perp \textbf{c}_R$ polarization. Features are plotted with an artificially narrow $\Gamma_f$=0.1 eV width for visual clarity.}
\label{fig:RIXSsimulation}
\end{figure*}

The vanadium L-edge polarization-dependent XAS spectra of VO$_2$ in the M1 phase are shown in Fig. 2(a, top), and are well studied in previous research \cite{20,21}. These high-resolution measurements were carried out using the total electron yield method at the elliptically polarized undulator beamline 4.0.2 of the Advanced Light Source, in the Vector Magnet endstation \cite{VMendstation}, and employ the same experimental geometry as the RIXS measurements. Separate spectral features associated with the $L_3$ and $L_2$ core hole symmetries are observed at $\sim$517 eV and $\sim$524 eV respectively, each having weak leading edge features followed by a strong high energy peak. Intensity near the $L_3$ maximum is greatest with polarization parallel to the axis of vanadium dimerization ($c_R$-axis), while intensity at $L_2$ is enhanced when polarization is normal to the dimer axis. The atomic multiplet simulation in Fig. 2(a, bottom) reproduces these characteristics, with a $\sim$0.5 eV discrepancy in some feature energies within the $\textbf{E} \parallel \textbf{c}_R$ channel.

%Previous studies have found that the detailed correspondence between experiment and theory for XAS can be improved by considering just a single vanadium atom hybridized with oxygen, and by treating the ground state as a non-coherent superposition of different occupied orbitals. The present study excludes oxygen hybridization to instead incorporate hybridization between vanadium dimers. The ground state is obtained as a pure state via diagonalization of the dimerized multiplet, for the sake of self-consistency in treatment of the ground state and RIXS final states.

%***work on the following:

Low energy excitations from 0-4 eV are measured by RIXS in Fig. 2(b). Broad features centered at roughly $\sim$0.6 eV and $\sim$2.3 eV are consistent with the energy gaps expected for excitation of a $d_{\parallel}$ electron into the octahedral $\pi^*$ ($t_{2g}$) and $\sigma^*$ ($e_g$) symmetry state manifolds, and have been labeled accordingly. Spectral intensity at low temperature is greatly suppressed within the $\Delta$=0.6 eV insulating gap of VO$_2$ \cite{4,OxygenKdbstar}, and is partly filled-in when the sample is heated into the metallic phase. Remarkably, the $\pi^*$ ($t_{2g}$) excitation feature has significant polarization dependence, and has an intensity maximum \emph{inside} the 0.6 eV insulating gap when measured at low temperature with polarization parallel to the $c_R$-axis. It is also noteworthy that upon cooling into the insulating phase, no feature appears at $\sim$1.4 eV, the expected energy of a $d_{\parallel}^s$ to $d_{\parallel}^*$ excitation in the itinerant limit ($1.4$ eV is roughly twice the inter-dimer $d_\parallel$ hopping parameter \cite{15}).

\begin{figure*}
\centering
\includegraphics[width = 16cm]{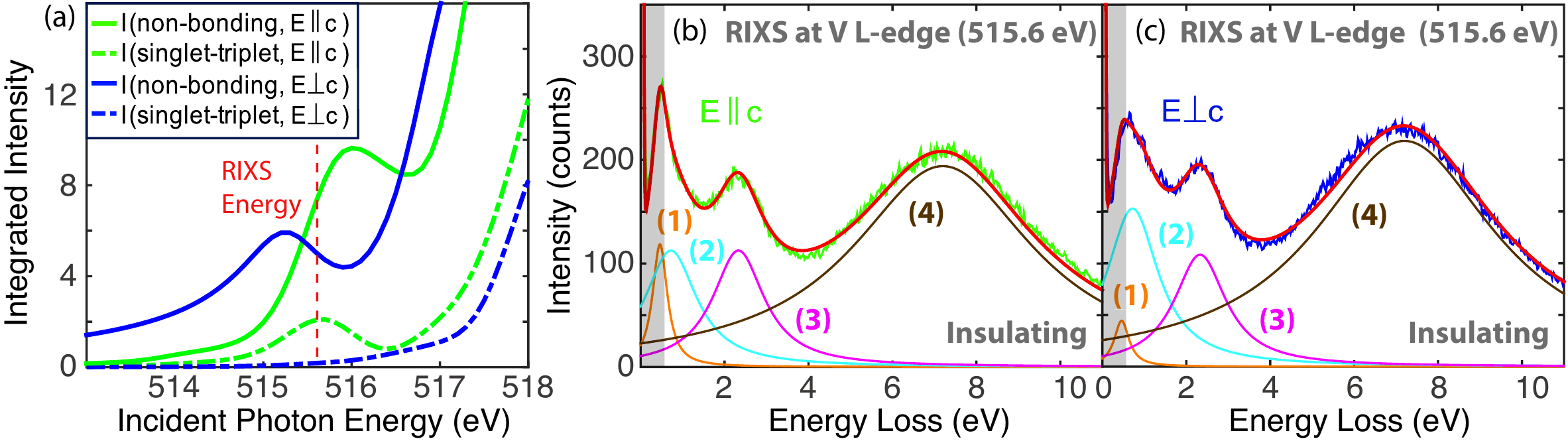}
\caption{{\bf{Polarization dependence and feature attribution}}: (a) Incident energy dependence of RIXS intensity in the singlet-triplet excitation and non-bonding $\pi^*$ ($t_{2g}$) excitation is simulated for the experimental conditions. The singlet-triplet excitation is best viewed with $\textbf{E} \parallel \textbf{c}_R$ polarization in the vicinity of the RIXS incident energy. (b-c) Four-Lorentzian fits of the experimental data measured with incident polarization (b) parallel and (c) perpendicular to the $c_R$-axis. Features nominally represent: (1) the singlet-triplet mode (energy E=0.46 eV, width $\Gamma$=0.43 eV); (2) $dd$ excitations into non-bonding $\pi^*$ ($t_{2g}$) orbitals (E=0.72 eV, $\Gamma$=1.5 eV); (3) $dd$ excitations into $\sigma^*$ ($e_{g}$) orbitals (E=2.33 eV, $\Gamma$=1.5 eV); and (4) metal-ligand charge transfer excitations (E=7.2 eV, $\Gamma$=5.25 eV). The singlet-triplet excitation constitutes 23\% weight of the combined feature (1-2) peak when polarization is parallel to the $c_R$-axis, and just 8\% under orthogonal polarization.}
\label{fig:fitting}
\end{figure*}

%Optical properties of correlated materials: Generalized Peierls approach and its application to VO2
%***this is too boring a transition

To understand these spectra, low energy excited states of the dimerized $V_2$ atomic multiplet model are used to calculate the RIXS spectral function in Fig. \ref{fig:RIXSsimulation}, via the Kramers-Heisenberg equation:
\begin{align*}
I(E,h\nu)=\sum\limits_f\sum\limits_g\left|\sum\limits_m\frac{\braket{f|T^{\dagger}|m}\braket{m|T|g}}{h\nu-E_{m}+E_{g}+i\Gamma_{m}/2}\right|^{2}\\
    \times\frac{\Gamma_{f}/2\pi}{\left(E-E_{f}\right)^{2}+\left(\varGamma_{f}/2\right)^{2}}
\end{align*}

Here, the spectral intensity is dependent on both the excitation energy ($E$) and the incident energy ($h\nu$), which determine the degree of resonance for scattering paths from the ground state 'g' to intermediate core hole states 'm' and the final excited states 'f', which are broadened by inverse lifetime terms ($\Gamma$). All final excited states fall within symmetric ($Q=0$) or antisymmetric ($Q=\pi$) sectors, and appear with different matrix elements for polarization polarization parallel and perpendicular to the dimer axis (see Fig. 3(a-b,d-e)). In the experimental geometry chosen for this study, the momentum (Q) transferred from the scattering event has a component of $Q=0.26\pi$ along the dimer axis, in units of the inverse distance between nearest neighbor vanadium atoms. This resulting RIXS spectrum is derived 84\% ($cos(0.26\pi/2)^2=0.84$) from the $Q = 0$ final state sector and 16\% ($sin(0.26\pi/2)^2=0.16$) from the $Q = \pi$ sector.

All of the symmetry sectors show qualitatively similar RIXS spectra, with prominent peaks at E$\sim$0.8 eV and E$\sim$2.3 eV representing transitions from the $d_{\parallel}$ orbital to non-bonding excited states involving the octahedral $\pi^*$ ($t_{2g}$) and $\sigma^*$ ($e_{g}$) manifolds, respectively. The $Q=\pi$ excitation sector differs from the optically accessible $Q=0$ spectrum in that a singlet-triplet excitation of the dimer is found at 0.42 eV, 26\% reduced from the singlet-triplet gap ($d_{\parallel}^s$ to $d_{\parallel}^t$ transition) expected from perturbation theory in the strongly correlated limit ($\frac{4t_\parallel^2}{U}$=4$\times$-0.8 eV$^2$/4.5 eV=0.57 eV). Charge transfer excitations between the vanadium atoms are not seen, as they occur at a higher energy scale than the plotted range. Calculated spectra for the experimental momentum value are shown in Fig. 3(c,f), and show the three features outlined in the ``Singlet" scenario of Fig. 1(b).

Close comparison between experiment and theory is complicated by the fact that most experimental features appear at energies larger than the band gap, and may be significantly broadened due to rapid decay of multiplet states into delocalized band excitations \cite{WrayFrontiers}. Plotting the experimental results over a larger energy range in Fig. 4(b-c) reveals that higher energy line features are qualitatively broader, and sharp line shapes comparable with experimental resolution are only found within the insulating gap. Taking this trend into account, the experimental data under each polarization condition are well fitted by four Lorentzians representing the three low energy features found in the simulation, as well as one high energy excitation at 7.2 eV, which can be principally attributed to metal-ligand charge transfer. Feature energies below E$<$4 eV are lower than nearby peak energies seen in the imaginary part of the dielectric constant by optical spectroscopies \cite{opticalFilmTiO2,optical1,optical2,optical3,optical4}. This is in part due to differing excitation symmetries \cite{BiermannOptical}, and can also be attributed to the spatially localized nature of the RIXS scattering process, which couples to coherent atomic-exciton-like final states. The direct RIXS excitations of strongly correlated systems with a single electron degree of freedom per atom (e.g. cuprates, vanadates) correspond closely with the difference between orbital site energies set by the crystal field \cite{SalaCuprateOrbiton}, whereas optical excitations represent transitions between band continua.

A particularly dramatic slope is seen from 0.2-0.5 eV under $\textbf{E} \parallel \textbf{c}_R$ polarization (Fig. 4(b)) and means that, given the above constraints, an good fit for that polarization condition must include a larger component of the sharp in-gap singlet-triplet mode. Attribution of this mode to the singlet-triplet dimer-breaking excitation is further supported by the fact that this sharp feature is no longer evident upon heating into the non-dimerized metallic phase (Fig. 1(b)), and that no analogous feature is seen in optical ($Q\sim 0$) measurements on analogous thin film samples \cite{opticalFilmTiO2}. With this feature attribution and better separated singlet-triplet and non-bonding modes, we anticipate that it would be possible to identify a more detailed line shape dressed by the interplay of coherent phonon states and the inter-vanadium hopping parameter as atomic positions relax following the breaking of the singlet bond.

To assess the likely accuracy of this fit, a summary of the calculated intensity of scattering in the singlet-triplet excitation (E$\sim$0.45 eV) versus non-bonding $\pi^*$ ($t_{2g}$) symmetries (E$\sim$0.8 eV) near the experimental RIXS incident energy of $h\nu$=515.6 eV is shown in Fig. 4(a). Within a $\pm$0.25 eV neighborhood surrounding the RIXS energy, the singlet-triplet feature accounts for 19.3\% of the intensity in a combined $t_{2g}$ feature under $\textbf{E} \parallel \textbf{c}_R$ polarization and 4.5\% of the intensity with $\textbf{E} \perp \textbf{c}_R$. These numbers show a good qualitative correspondence with values of 23\% and 8\% respectively from the fit. The model appears to underestimate the overall intensity of such a combined feature in the $\textbf{E} \perp \textbf{c}_R$ channel, however the calculated non-bonding resonance peaks near the RIXS energy in both polarization channels have a close $\pm$0.2 eV correspondence with features seen by XAS (Fig. 1(a, top)).

In summary, RIXS has been used to measure the energies of single particle transitions between the significant orbital manifolds of VO$_2$. By comparison with a first principles-derived numerical model, we find a strong correspondence between this spectrum and the electronic states expected in a strongly correlated picture for low temperature vanadium dimers. Scattering matrix elements are found to enable the first experimental measurement of the singlet-triplet excitation that breaks the singlet spin bond of a vanadium dimer. Polarization and temperature dependence in the experimental spectra are used to identify a strong candidate for this feature at E=0.46 eV. These results provide a window into the gap structure and high energy landscape underlying the metal-insulator transition of VO$_2$, and more generally demonstrate the power of the RIXS technique as an incisive probe of correlated energetics in transition metal compounds.

\textbf{Acknowledgements:} We are grateful for discussions with G. Kotliar. RIXS measurements were performed at the ADRESS beamline of the Swiss Light Source using the SAXES instrument jointly built by Paul Scherrer Institut, Switzerland and Politecnico di Milano, Italy. Work at NYU was supported by the MRSEC Program of the National Science Foundation under Award Number DMR-1420073. Research at Stanford was supported through the Stanford Institute for Materials and Energy Sciences (SIMES) under contract DE-AC02-76SF00515 and the LCLS by the US Department of Energy, Office of Basic Energy Sciences. The Advanced Light Source is supported by the Director, Office of Science, Office of Basic Energy Sciences, US Department of Energy under Contract No. DE-AC02-05CH11231. J.P. and T.S. acknowledge financial support through the Dysenos AG by Kabelwerke Brugg AG Holding, Fachhochschule Nordwestschweiz, and the Paul Scherrer Institut.

\end{document}